\documentstyle[a4,11pt,epsf]{article}
\epsfverbosetrue
\font \math=msbm10 scaled \magstep 0

\begin{document}

\begin{titlepage}
 



\vspace{5mm}
 
\begin{center}
{\Huge Critical Behaviour of the Two Dimensional\\ [3mm]
 Step Model}\\[15mm]
{\bf A.C. Irving and R. Kenna\footnote{Supported by EU Human Capital 
and Mobility Scheme Project No. CHBI--CT94--1125} \\
DAMTP, University of Liverpool L69 3BX, England} \\[3mm]
August 1995
\end{center}
\begin{abstract}
We use finite--size scaling of     Lee--Yang partition function zeroes to
study the critical behaviour of the two dimensional step  or  sgn  $O(2)$ 
model. We present evidence that, like the  closely  related  $XY$--model,
this has a phase transition from a disordered high  temperature  phase to 
a low temperature massless phase where the  model  remains critical.  The 
critical  parameters  (including  logarithmic corrections) are compatible 
with those of the $XY$--model  indicating  that both models belong to the 
same universality class.
\end{abstract}


\end{titlepage}

\section{Introduction}
\label{sec:intr}
In a recent paper \cite{KeIr95}, we demonstrated the power of finite--size 
scaling applied to Lee--Yang zeroes \cite{LY} in uncovering logarithmic
corrections to scaling in the two--dimensional $XY$-- or $O(2)$ spin model.
In this paper we apply the same techniques to the closely related \lq step 
model\rq{} \cite{GuJoTh72,GuJo73}, 
also known \cite{LeSh87,LeSh88} as the sgn $O(2)$ 
model. The question of criticality of this model has, until now, been unresolved 
despite several analyses based on high temperature series (see \cite{LeSh88} 
for a review) and on numerical simulation \cite{NyIr86,SVWi88}. The interest 
in the model arises from its possible membership of the $XY$--model universality 
class which exhibits the Kosterlitz--Thouless [KT] phase transition \cite{KT}. 
Like the $XY$--model, the step model has a 
configuration space which is globally and continuously symmetric.
Unlike the $XY$--model, however, the interaction function 
is discontinuous and the Mermin--Wagner theorem \cite{MeWa66} does not apply. 
Nonetheless, it is expected that if a phase transition exists in the step model, 
it should not be to a phase with long range order \cite{GuJo73,SVWi88,GuNy78}.

S\'anchez--Velasco and Wills \cite{SVWi88} presented evidence of critical
behaviour starting at $\beta_c=1/T_c=0.91\pm0.04$. This was based on 
finite--size scaling [FSS] of the spin susceptibility. Since the associated 
critical index $\eta(T_c)$ was significantly greater than that measured for 
the $XY$--model, it was concluded that the step and $XY$ models are not 
in the same universality class. In this paper we present evidence that the 
step model is {\em not} critical at that temperature. However it {\em is} 
critical at lower temperatures with a critical index $\eta(T_c)$ compatible 
with the $XY$ value. The accuracy afforded by the Lee--Yang zeroes study 
is a crucial part of the analysis.

\section{The step model and the $XY$--model}
\label{sec:step}
Consider the  partition  function 
\begin{equation}
 Z(\beta,h)
 =
 \sum_{\{{\vec{s}}_x\}}{
                   e^{- \beta H
                    + h {\hat{n}}\cdot{\vec{M}}
                   }
               }
\, ,
\label{eqn:pf}
\end{equation}
where the Boltzmann factor is $\beta = 1/kT$, ${\hat{n}}$ is a unit vector 
defining the direction of the external magnetic field and $h$ is a scalar 
parameter representing its strength. The summation is over all 
configurations open to the system and ${\vec{s}}_x$ is a unit length 
two-component spin at each site $x$ in the cubic lattice 
$\Lambda\equiv L^d$ ($d=2$). The magnetisation for a given configuration is
\begin{displaymath}
 {\vec{M}} =
 \sum_{x \in \Lambda}
 {\vec{s}}_x
\, .
\end{displaymath}
In the case of the $XY$--model, the interaction hamiltonian is 
\begin{displaymath}
 H_{{XY}}
 =
 -
 \sum_{x \in \Lambda}          
 \sum_{\mu = 1}^{d}
 {\vec{s}}_x\cdot  
 {\vec{s}}_{x+\mu}
\, ,
\end{displaymath}
while for the step model it is
\begin{displaymath}
 H_{\rm{step}}
 =
 -
 \sum_{x \in \Lambda}
 \sum_{\mu = 1}^{d} \hbox{sgn}(
 {\vec{s}}_x\cdot
 {\vec{s}}_{x+\mu})
\, .
\end{displaymath}
Thus the continuous (cosine) dependence of the interaction energy
in the usual $XY$--model is replaced by a discrete step function dependence.

The leading infinite volume critical behaviour of the 2D $XY$--model is 
characterized by exponential divergences (essential singularities) in the 
thermodynamic functions \cite{KT}. In terms of the reduced  temperature 
$t\equiv 1-\beta/\beta_c \rightarrow 0^+$ the (leading) 
infinite volume scaling behaviour 
of the correlation length and the zero-field magnetic susceptibility is 
(respectively) \cite{KT}
\begin{eqnarray}
 \xi_\infty(t) & \sim & e^{at^{-\nu}} \, , \label{eqn:ktxi}
 \\
 \chi_\infty(t) & \sim & \xi_\infty^{2-\eta}\, , \label{eqn:ktchi}
\end{eqnarray}
where $\nu = 1/2$ and $\eta = 1/4$. The $XY$--model remains critical for all 
$ \beta > \beta_c$.

For models obeying the Lee--Yang theorem \cite{LY}, 
the partition function zeroes in the magnetic
field strength ($h$) plane (the Lee--Yang zeroes) 
are all on the imaginary axis. In the high temperature 
phase these zeroes remain away from the
real axis, pinching it only as $t \rightarrow  0^+$ (in the 
thermodynamic limit).
The zero  closest to the real axis marks the edge of the distribution 
of zeroes and is known as the Yang--Lee edge, $z_{\rm{YL}}$.
The theorem has been proved only for certain models, the $XY$--model included 
\cite{DuNe75}.
In \cite{KeIr95} we used this fact to show the above leading
critical behaviour for the $XY$--model and 
the corresponding behaviour of the Yang--Lee edge, $z_{\rm{YL}}$ are, in 
fact, modified by logarithmic corrections:
\begin{eqnarray}
 \chi_\infty (t) & \sim & \xi_\infty^{2 - \eta} t^{r}\, ,
\label{eqn:chis}
 \\
  z_{\rm{YL}}(t)   &  \sim & \xi_\infty^{\lambda} t^{p}\, ,  
\label{eqn:edges}
\end{eqnarray}
where 
\begin{equation}
\lambda= -\frac{1}{2}(d+2 -\eta)  = -{{15}\over 8}
\label{eqn:lambda}
\end{equation}
and the parameters $r$ and $p$ ($=-r/2)$ are logarithmic correction indices 
\cite{KeIr95}. The corresponding FSS behaviour for the susceptibility and 
the first zero $z_1$ ($=z_{YL}$) at $t=0$ is \cite{KeIr95}
\begin{eqnarray}
 \chi_L (0) & \sim & L^{2 - \eta} (\ln{L})^{-\frac{r}{\nu}}\, ,
\label{eqn:chifss}
 \\
  z_1(0)   &  \sim & L^{\lambda}  (\ln{L})^{r}  
\label{eqn:edgefss}
\, .
\end{eqnarray}
For the 2D $XY$--model the numerical value of $r$ was found to be small 
but non-zero ($-0.023\pm 0.010$) \cite{KeIr95}.

The objects of the present analaysis were to establish
\begin{enumerate}
\item if the scaling behaviour of the very precisely determined Yang--Lee 
edge would unequivocably determine whether the step model had a critical 
phase 
\item if so, whether the phase transition is in the same universality class 
as the $XY$--model.
\end{enumerate}

\section{Method and results}
\label{sec:meth}
The methods used are those described in \cite{KeIr95}. A single cluster 
algorithm \cite{Wo89} is used to generate a large number of measurements 
(100K for each lattice size $L$ and temperature $1/\beta$) of the energy 
$H$ and the magnetisation ${\vec{M}}$ at zero external magnetic field. 
Such data were obtained for lattice sizes $L=32$, $48$, $64$, $128$ and 
$256$ covering the $\beta$ range $0.86$ to $1.40$ with  varying degrees 
of spacing. Multi--histogram techniques \cite{FeSw88,KaKa91} were used to 
combine data at different values of $\beta$ and so obtain detailed $\beta$ 
dependence. In the neighborhood of possible critical points we used 
sufficiently fine spacing (typically 0.025) to ensure adequate overlap 
between histograms for a given size of lattice.

The partition function for a complex magnetic field ($h=h_r+ih_i$,
$h_r,h_i \in \mbox{\math R} $) can be written in terms of a real and an imaginary 
part \cite{KeIr95,KeLa93,KeLa94} as
\begin{equation}
  Z(\beta,h_r + i h_i)
  =
  {\rm{Re}} Z(\beta,h_r + i h_i)
  + i
  {\rm{Im}} Z(\beta,h_r + i h_i)
\, ,
\end{equation}
where
\begin{eqnarray}
 {\rm{Re}} Z(\beta,h_r + i h_i)
 & = &
 Z(\beta,h_r)
 \langle
 \cos{(h_i M)}
 \rangle_{\beta,h_r}
\label{ReZ}
\, ,
\\
 {\rm{Im}} Z(\beta,h_r + i h_i)
 & = &
 Z(\beta,h_r)
 \langle
 \sin{(h_i M)}
 \rangle_{\beta,h_r}
\label{ImZ}
\, .
\end{eqnarray}
Here the subscripts indicate that the expectation values are taken at
(inverse) temperature $\beta$ and in a (real) external field $h_r$. 

No specific proof of the Lee--Yang theorem exists for the step model.
We can, however, use (\ref{ReZ}) and (\ref{ImZ}) to (numerically)
determine the loci along which the real and imaginary parts of the 
partition function separately vanish. These formulae concern
expectation values of real quantities in a real external field
and at no stage is a simulation involving a complex action involved.
The Lee--Yang zeroes are then the points in the complex $h$--plane
where the loci intersect \cite{KeLa94}. 
We have determined these loci 
and thereby the Lee--Yang zeroes close to the real $h$--axis. 
We were able to determine the first 15 zeroes and found
that they lie on the imaginary $h$--axis for all the lattices 
studied. Thus we have numerical evidence that the step model
obeys the Lee--Yang theorem. The following analysis applies to
the first zeroes only (the Yang--Lee edge) and we defer the study of higher 
zeroes to a later paper \cite{bbig}.

\begin{figure}[htb]
\vspace{9.8cm}
\includegraphics{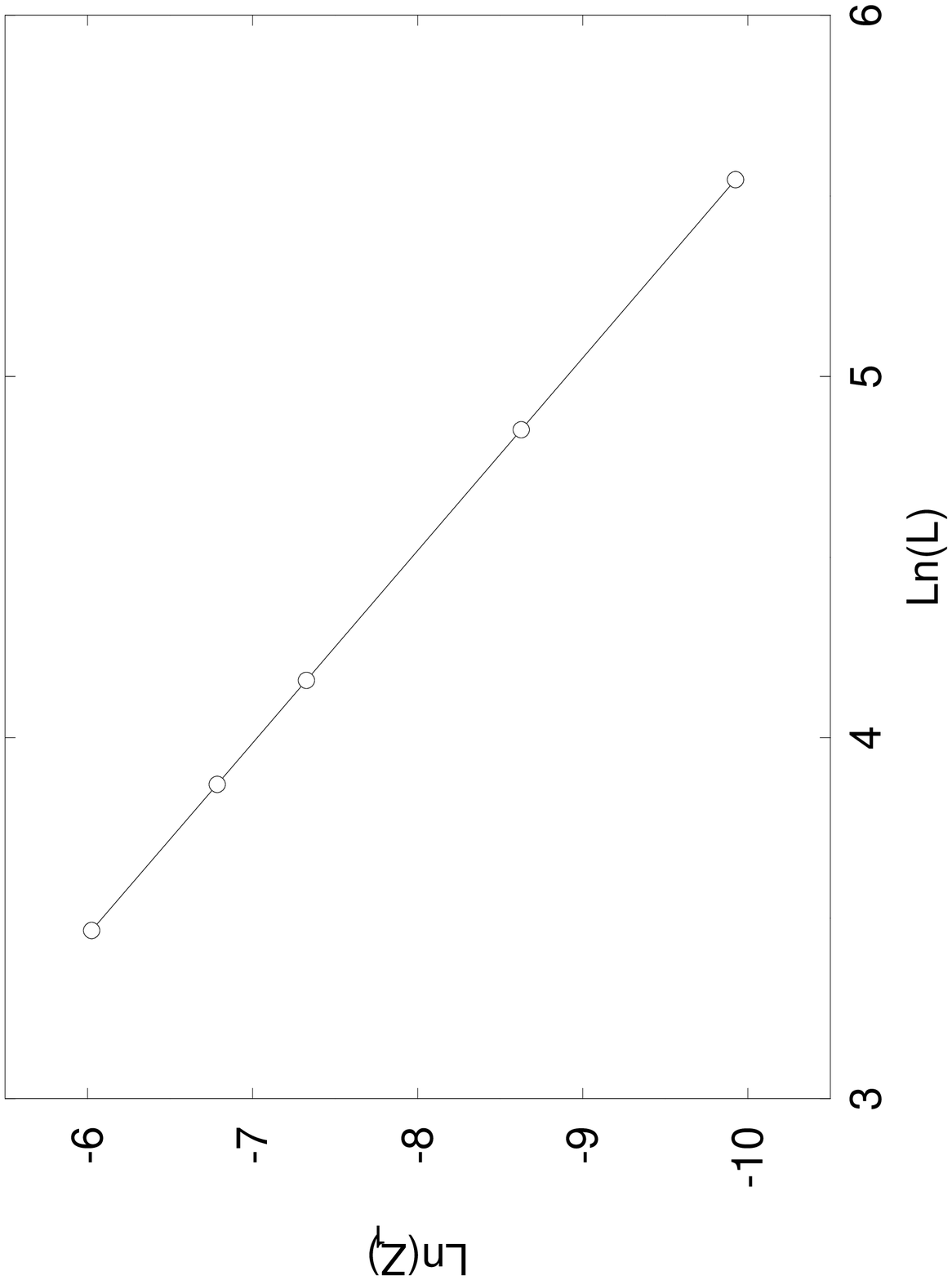}
\caption{Scaling behaviour of the Yang--Lee edge at a typical
critical $\beta$ value ($\beta = 1.22$).
}
\label{fig:typscal}
\end{figure}

The analysis began with a rough search for the leading critical behaviour 
predicted by (\ref{eqn:edgefss}). An independent test was also made 
using the (less accurate) susceptibility data and  (\ref{eqn:chifss}). 
Both methods indicated critical behaviour setting in for $\beta$ 
\raisebox{-.75ex}{ {\small \shortstack{$>$ \\ $\sim$}} } $1.2$. 
In Fig.~\ref{fig:typscal} we show a typical log--log plot of 
$z_1$ the Yang--Lee edge versus $L$. This is at a typical candidate value 
of the critical temperature ($\beta=1.22$). The errors are considerably 
smaller than the symbols. For example, at $\beta=1.22$ we found 
$z_1=0.0024136(7)$ and $0.00017902(6)$ at $L=32$ and $128$ respectively. 
The slope of Fig.~\ref{fig:typscal} gives the effective leading index 
$\lambda_{\hbox{eff}}$ ignoring corrections. At $\beta=1.22$ this is 
$\lambda_{\hbox{eff}} = -1.8761(2)$ where the chi-squared per degree
of freedom ($\chi^2/{\rm{dof}}$) for the linear fit shown is $0.85$. 
In Fig.~2(a) we display the result of such fits as a function of $\beta$. 
The effective exponent $\lambda_{\hbox{eff}}$ is just the slope of the 
log--log linear fit which should obtain if critical behaviour is present 
(ignoring logarithmic corrections). The corresponding $\chi^2/{\rm{dof}}$ 
for the linear fit is also shown. Acceptable values are only found for 
$\beta$ in excess of around 1.2. To quantify this statement we demand 
\begin{equation}
\chi^2/{\rm{dof}} \leq 2.0
\label{eqn:chisq}
\end{equation}
which means 
$\beta \raisebox{-.75ex}{ {\small \shortstack{$>$ \\ $\sim$}} } 1.185$.
We note that the corresponding  values of $\lambda_{\hbox{eff}}$ 
($ \raisebox{-.75ex}{ {\small \shortstack{$<$ \\ $\sim$}} } -1.872(2)$) 
include that ($-15/8 = -1.875$) corresponding to the KT prediction.

\begin{figure}[htb]
\vspace{13cm}
\includegraphics{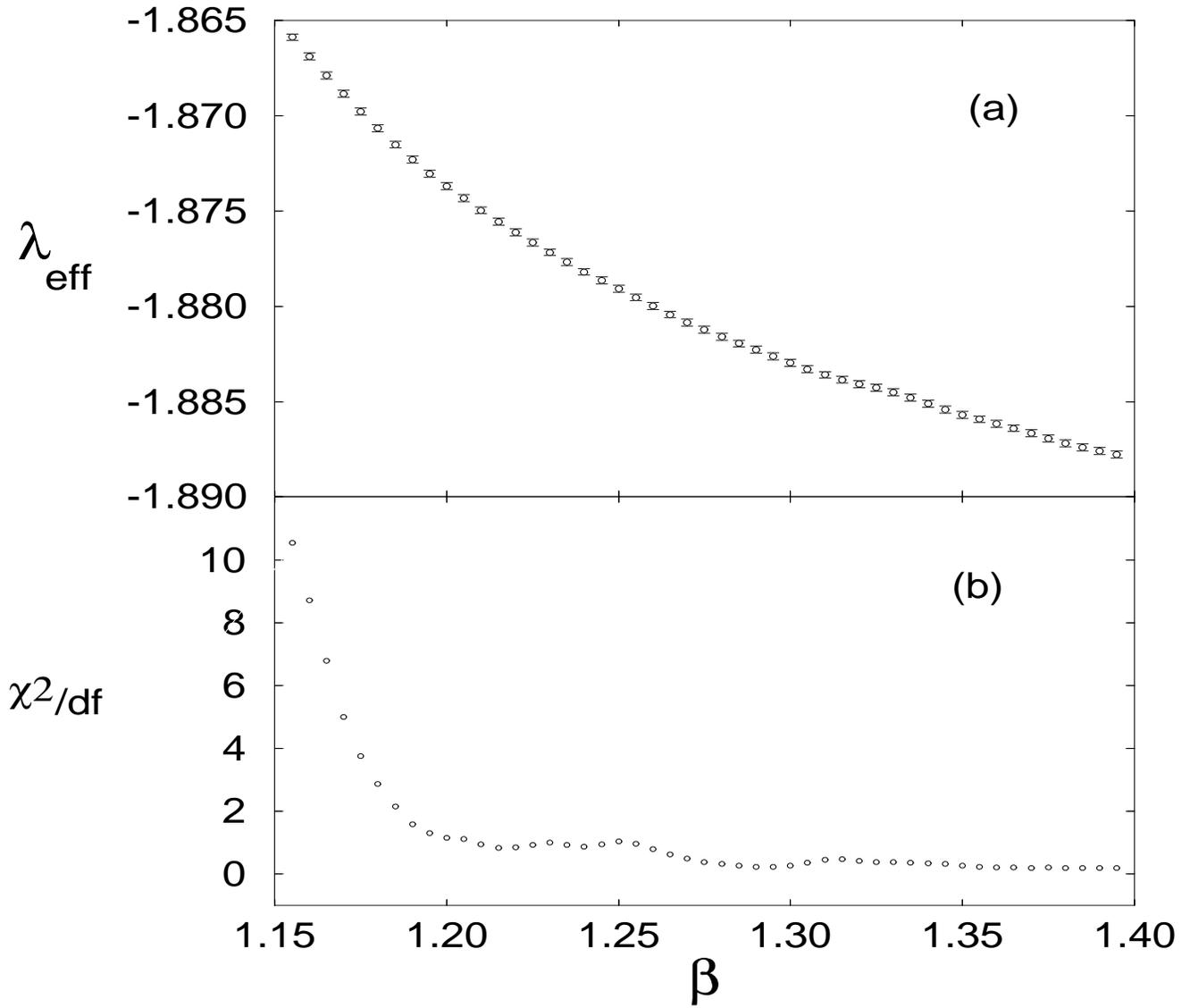}
\caption{Test of expected leading critical behaviour: 
(a) the effective exponent $\lambda_{\rm{eff}}$ 
(slope of a straight line fit)
and (b) the corresponding $\chi^2/{\rm{dof}}$ versus $\beta$.}
\label{fig:lambda}
\end{figure}

Fig.~\ref{fig:lambda} is evidence for ({\em{i}}) the validity of FSS 
over a range of values $\beta \ge \beta_c \simeq 1.185$ and ({\em{ii}}) 
at $\beta_c$ the exponent
$\lambda_{\hbox{eff}}$ very close to the expected KT value (-15/8).
Observation ({\em{i}}) means that, as in the $XY$ case, the system 
remains critical for 
$\beta \ge \beta_c$ and ({\em{ii}}) is evidence that
$\lambda$ is in fact -15/8 and the question of whether or not the step
model belongs to the same universality class as the $XY$ model must now
be answered by determination of the correction exponent $r$. 

We therefore {\em assume} the behaviour (\ref{eqn:edgefss}) with
$\lambda = -15/8$  at $\beta=\beta_c$. The expected 
leading behaviour is removed and linear fits to
\begin{equation}
\ln \left( z_1 L^{15/8}\right) \quad\hbox{vs.}\quad\ln\ln L
\end{equation}
performed.  The  results are shown in Fig.~3. Since the value $\lambda=-15/8$ 
($\eta=1/4$) is only expected at $\beta_c$ these results can be used to 
identify the possible values of critical temperature and to test for the 
presence of logarithmic corrections as in the $XY$--model \cite{KeIr95}.

Applying the same criterion (\ref{eqn:chisq}) as for the leading behaviour, 
we search for a range of $\beta_c$ values giving an acceptable fit. We find
\begin{equation}
1.195\leq\beta_c\leq 1.295\quad\hbox{and correspondingly,}\quad
0.009\geq r \geq -0.034\, .
\end{equation}
The range of acceptable $r$ values includes that found \cite{KeIr95} for 
the $XY$--model ($-0.023 \pm 0.010$) and that corresponding to no 
logarithmic corrections ($r=0$). As in our previous work \cite{KeIr95}, 
this range excludes the  prediction $r = -1/16$ coming from an 
approximate renormalisation group treatment of the $XY$--model \cite{KT}. 
Thus we conclude that the present data are compatible with the step model 
being in the same universality class as the $XY$--model. We do not, however,
exclude other possibilities.

The susceptibility data are consistent with the above analysis. If one
assumes the KT value of $\eta(\beta_c)=1/4$, one can construct a
so-called  Roomany--Wyld beta function approximant \cite{RW} from the
finite--size data and use
its zero to locate $\beta_c$ \cite{KeIr95}. These approximants, based on
pairs of lattice size $L, L'$, converge
very rapidly \cite{RW}. We estimate $\beta_c=1.22 \pm 0.02$. 
This last analysis
of course neglects possible logarithmic corrections to scaling. 

We have also studied the specific heat. As for the $XY$--model, the step
model data show a broad peak with no obvious relationship to the
position of the leading critical point. The finite--size dependence is
not dramatic and is likely to be of little value in further elucidating the 
criticial behaviour. A related question is to what extent one can make
use of the Fisher zeroes \cite{KeLa93,Fi72}, 
i.e. zeroes of the partition function in the
complex $\beta$ plane at zero external magnetic field $h$. For both this
and the $XY$--model, these are much harder to locate than the Lee--Yang
zeroes and are consequently less accurately determined.

\section{Conclusions}
\label{sec:conc}
The use of finite--size scaling applied to Lee--Yang zeroes has allowed
us to present detailed evidence of critical behaviour in the
two-dimensional step (sgn $O(2)$ spin) model. 
The data are consistent with this model being in the same universality
class as the $XY$-- ( $O(2)$ spin) model. That is, it undergoes a 
Kosterlitz--Thouless type transition with susceptibility index 
$\eta({\beta_c})=1/4$ and we determine that $\beta_c$ lies in the range
$1.195 \leq \beta_c \leq 1.295$. 
With the available statistics, we found the logarithmic correction exponent
to lie in the range $ -0.034 \leq r \leq 0.009$.
This should be compared with our measurement for the $XY$ correction exponent
\cite{KeIr95}, $-0.033 \leq r \leq -0.013$ with which it is compatable.
The step model results are however also compatable with no logarithmic corrections 
($r=0$ corresponds to $\beta_c = 1.21$ and $\chi^2/{\rm{dof}}= 0.92$).

The Mermin--Wagner theorem \cite{MeWa66} does not apply directly 
to the step model because of the discontinuous nature of the interaction
hamiltonian.
However, it has long been believed \cite{GuJo73,SVWi88} that if 
a phase transition does exist, it will not involve a phase with long range
order. The evidence presented here supports this view.

This raises a question as to the nature of the 
mechanism driving the phase transition 
in the step model. 
The KT phase transition of the
$XY$--model is understood to be driven by the binding/unbinding of topological
solutions (vortices). 
However, the energetics of vortex formation are
very different in the step model \cite{GuNy78,LeSh88,SVWi88}. 
Since vortices with effectively zero excitation energy
can be created at all non-zero temperatures,
the usual KT argument does not naturally lead one to expect 
such a phase transition in the step model.

If this is indeed the case, some other driving mechanism must be 
responsible for any phase transition. It would then be remarkable 
if --- as the evidence presented here indicates -- both models 
belong to the one universality class.


\newpage

\newpage

\begin{figure}[htb]
\vspace{13cm}
\includegraphics{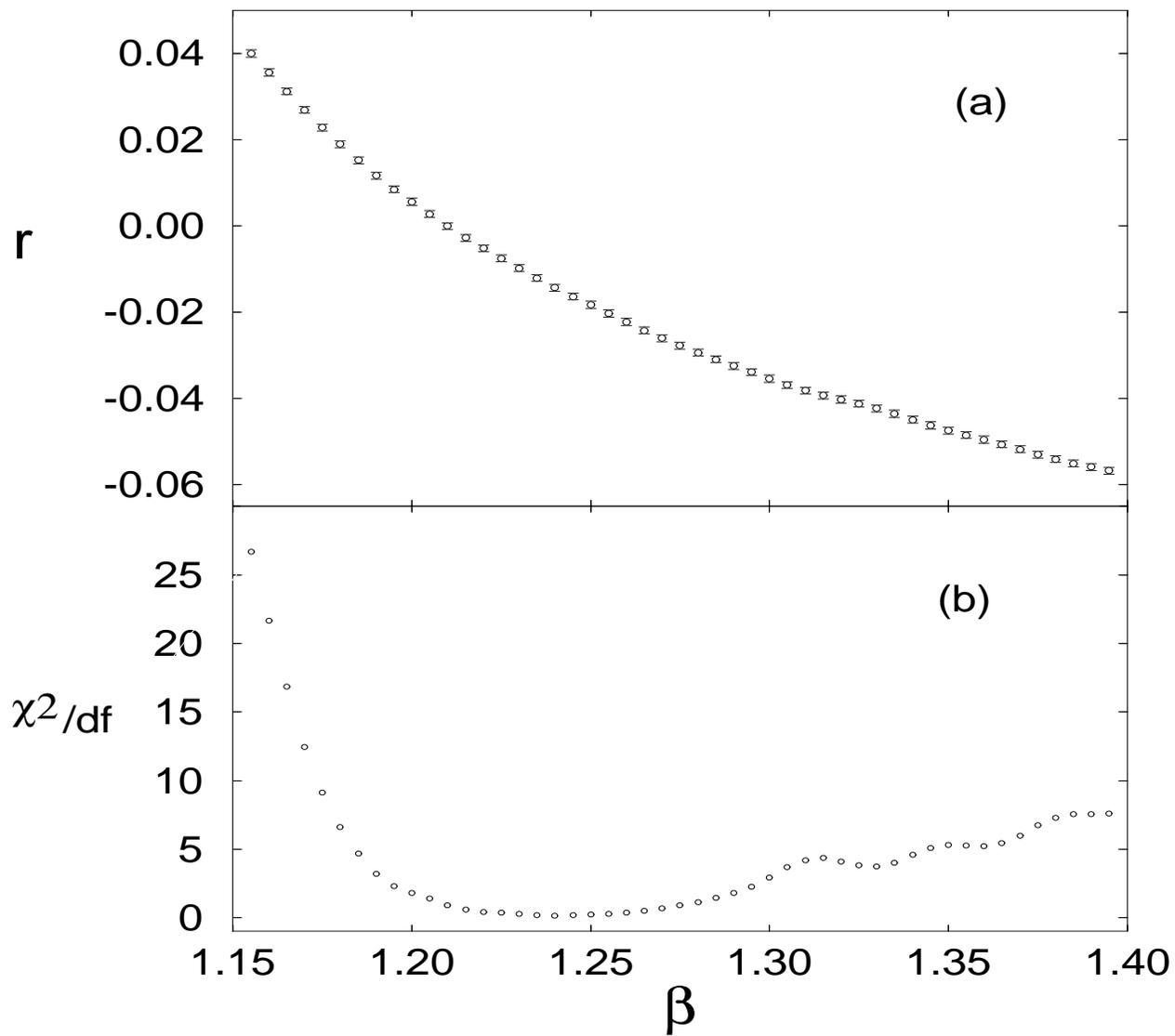}
\caption{Logarithmic corrections: (a) the logarithmic correction
exponent $r$ to the Yang--Lee edge is shown as a function of the assumed
critical coupling $\beta_c$ and (b) the corresponding $\chi^2/{\rm{dof}}$.
}
\label{fig:r}
\end{figure}

\end{document}